# Tuning degree distributions of scale-free networks

C.C. Leary[1,2], M. Schwehm[*,1], M. Eichner[1], H.P. Duerr[1]

[1] Department of Medical Biometry, University of Tübingen, Westbahnhofstraße 55, 72070 Tübingen, Germany

[2] Department of Mathematics, State University of New York at Geneseo, 1 College Circle, Geneseo, NY 14454, USA

[*] Corresponding author: schwehm@informatik.uni-tuebingen.de

## ABSTRACT

Scale-free networks are characterized by a degree distribution with power-law behavior and have been shown to arise in many areas, ranging from the World Wide Web to transportation or social networks. Degree distributions of observed networks, however, often differ from the power-law type and data based investigations require modifications of the typical scale-free network.

We present an algorithm that generates networks in which the skewness of the degree distribution is tuneable by modifying the preferential attachment step of the Barabási-Albert construction algorithm. Skewness is linearly correlated with the maximal degree of the network and, therefore, adequately represents the influence of superspreaders or hubs. By combining our algorithm with work of Holme and Kim, we show how to generate networks with skewness $\gamma$ and clustering coefficient $\kappa$, over a wide range of values.

# I. INTRODUCTION

Since the development of effective computational tools, intense research has broadened our understanding of the structure of networks. Theoretical and observational work have pointed out differences and commonalities among networks ranging from the World Wide Web to transportation networks to social networks associated with sexually transmitted diseases [1-7]. The contact structure of the underlying network has been shown to be a critical feature in applications ranging from biology [8] to the internet [9-11] to sociology [12] to epidemiology [11, 13-17].

Scale-free contact networks, in which the probability that a randomly chosen vertex has degree $k$ is proportional to $k^{-\alpha}$, have attracted much attention. They appear frequently in real settings [18] and, thanks to their relative abundance of high-degree vertices, are particularly suited to the rapid transmission of things (ideas, diseases, viruses) across the network. Scale-free networks are also hardy—they are difficult to disconnect by random failure or attack [10, 11, 14, 16, 17, 19].

In 1999 Barabási and Albert published their preferential attachment algorithm, which generates scale-free networks [20]. Beyond merely generating these networks, the Barabási-Albert scheme produces networks in which vertices are added to the network serially, mimicking the growth of real networks over time. The "rich get richer" method of connecting new vertices to the network is simple, intuitive and plausible, and thus the algorithm has been widely studied and widely used by modelers.

Inevitably, the networks generated by the Barabási-Albert algorithm were found to not match every potentially important property of real scale-free networks. One such characteristic, introduced by Watts and Strogatz [1], is the clustering coefficient. For a vertex $v$ with $n$ neighbors, the clustering coefficient of $v$ is defined to be $\kappa_v = \frac{E_v}{{}_nC_2}$, where $E_v$ is the number of edges among the $n$ neighbors of vertex $v$ and ${}_nC_2 = \frac{n(n-1)}{2}$ is the maximum number of edges possible among those $n$ neighbors. Then $\kappa$, the mean clustering coefficient, is the average of $\kappa_v$ across all vertices $v$ in the network. The BA algorithm generates networks in which the clustering coefficient $\kappa$ is asymptotically 0, in contrast to the positive clustering observed in many real social and physical networks [21].

At the same time, technological, economic and social studies have uncovered many examples of naturally occurring networks that are decidedly not scale-free. (For a brief survey, see [18].) Large scale data-based simulations of contact patterns in connection with airborne disease transmission in Southeast Asia and in Vancouver have also generated degree distributions that differ significantly from the power law that characterizes scale-free networks [22-24].

In response to this situation, the suite of methods to generate networks has grown. Deterministic models have been explored in which both the degree distribution and the clustering coefficient can be altered [25-27]. Other authors have investigated the stochastic modeling of networks in which both the degree distribution and the clustering are tunable [28, 29]. The Barabási-Albert algorithm itself can be used to create scale-free networks with a range of mean clustering coefficients [30-34]. What has been missing is an algorithm that uses preferential attachment to generate networks whose degree distributions can deviate from scale-free.

In this paper we describe a new generalization of the Barabási-Albert approach. We use a tuned preferential attachment step as well as the triad formation step of Holme and Kim [32] in order to generate networks that grow over time in such a way that both the degree distribution and the clustering coefficient of the resulting network are tunable. The algorithm combines the intuitive attractiveness of the BA algorithm with enough flexibility to produce highly clustered networks with degree distributions that differ significantly from scale-free.

## II. METHODS

A contact network of 10,000 individuals was created using a generalization of the Holme and Kim [32] tuneable clustering version of the Barabási-Albert scale-free network generation algorithm. An initial collection of 8 vertices was connected in a cycle. After this, the remaining 9992 vertices were connected one at a time to the existing network. Vertex $v$ was attached by first choosing a vertex $w$ from the existing network using the preferential attachment scheme described below. Vertices $v$ and $w$ were connected (bidirectionally), and then $v$ was connected to three other vertices chosen independently by either a triad formation step (with probability $c$) or by the preferential attachment scheme (with probability $1-c$). Construction was completed by revisiting the initial eight vertices and connecting each of them to three other vertices either by triad formation or preferential attachment.

The preferential attachment algorithm used was a generalization of that described by Barabási and Albert [20]. In the classic BA algorithm, the probability that a vertex $w$ of degree $k$ is chosen to be attached to vertex $v$ is proportional to $k$. We altered this algorithm in the following way: The vertices $\{w_1, w_2, \ldots, w_n\}$ in the existing network were ordered by their degree from largest to smallest, and then a random number $r$ was chosen. The original BA preferential attachment algorithm chooses $r$ from a uniform distribution, but we alter this by using a tuning distribution with range [0,1] and expected value $\mu_T$, as described below. Vertex $v$ was then attached to vertex $w_j$ if

$$\frac{\sum_{i=1}^{j-1} \deg(w_i)}{\sum_{i=1}^{n} \deg(w_i)} \leq r < \frac{\sum_{i=1}^{j} \deg(w_i)}{\sum_{i=1}^{n} \deg(w_i)} .$$

The tuning distributions used for Figs. 1-4 were simple linear distributions, with probability density functions

$$f_m(x) = \begin{cases} mx + (1 - \frac{1}{2}m) & 0 \leq x \leq 1 \\ 0 & \text{otherwise} \end{cases},$$

with expectation $\mu_T = \frac{1}{2} + \frac{1}{12}m$. The tuning distribution $f_0$ with $m = 0$ and $\mu_T = \frac{1}{2}$ is the uniform distribution on the interval [0,1], and thus this parameter setting (with $c = 0$) corresponds to the original BA preferential attachment scheme. The tuning distribution used for Fig. 5 was the beta distribution given by

$$f(x) = \frac{(x - x_l)^{\alpha - 1}(x_u - x)^{\beta - 1}}{\mathrm{B}(\alpha, \beta)(x_u - x_l)^{\alpha + \beta - 1}},$$

with parameters $\alpha > 0$ and $\beta > 0$, in which $x_l$ and $x_u$ represents the lower and upper bound of the distribution, respectively (here: $x_l = 0$ and $x_u = 1$) and $\mathrm{B}(\alpha, \beta)$ denotes the beta function. The beta distribution offers more flexibility than the linear distributions and can produce almost all shapes, e.g. uniform ($\alpha = 1$ and $\beta = 1$), U-shaped ($\alpha < 1$ and $\beta < 1$), bell-shaped ($\alpha > 1$ and $\beta > 1$), right-skewed ($\alpha > 1$ and $\beta < 1$), or left-skewed ($\alpha < 1$ and $\beta > 1$).

The triad formation step of Holme and Kim [32], was used to induce clustering in the network. When vertex $v$ underwent a triad formation step, a vertex $w$ was chosen uniformly at random from among the set of vertices adjacent to $v$. Then a vertex $u$ was chosen uniformly from the neighbors of $w$ and vertex $v$ and vertex $u$ were connected (assuming that they were not already connected).

The algorithm was executed 100 times for each $\mu_T \in \{0.4, 0.45, 0.5, 0.55, 0.6\}$ and each $c \in \{0, 0.25, 0.5, 0.75, 1\}$, yielding 2500 total networks for the analysis. In order to test the sensitivity of our results to the initial configuration of vertices, a further 2500 networks were generated beginning with a complete graph on nine vertices (which maintains an average degree of 8 in the network).

## III. RESULTS

We report results from the networks in which the initial eight nodes were connected in a ring. The results for networks that were seeded with a complete graph were similar.

Independent of the parameter settings, the 10,000 vertices in the network had an average of 8 contacts. As the tuning distribution used in the preferential attachment steps of the network creation algorithm influenced the amount of preference gained by high-degree vertices, altering this distribution allowed us to influence the degree distribution of the resulting network, as illustrated in Fig. 1. In the following analysis (i) the subscript $T$ refers to a tuning distribution while a subscript $D$ refers to the degree distribution of a network, (ii) we report the skewness of a distribution as $\gamma^3$ and call its cube root $\gamma$ the root skewness of the distribution and (iii) we refer to the degree distribution generated by the classic scale-free algorithm of Barabási and Albert ($\mu_T = 1/2$, $c = 0$) as the scale-free (SF) degree distribution. Realizations of 100 SF degree distributions show a normally distributed root skewness with mean $\gamma_D = 2.26$. Varying the clustering probability does not substantially change the skewness (Fig. 2), e.g. even for the maximum value of the clustering parameter ($c$=1) the mean root skewness increases only to $\gamma_D = 2.29$.

### A. Overskewness & underskewness

We call a degree distribution overskewed if the skewness exceeds that of the SF degree distribution, and underskewed if the skewness falls short of this standard skewness. Overskewed degree distributions were produced by applying tuning distributions with expectation $\mu_T < 1/2$, which make connections to vertices with high degree more likely (Fig. 1, row 1 and Fig. 2). Similarly, underskewed degree distributions were produced by applying tuning distributions with expectation $\mu_T > 1/2$, which make connections to vertices with few contacts more likely (Fig. 1, row 3 and Fig. 2). Hence, the root skewness of the degree distribution is inversely related to the expectation of the tuning distribution. The slope of this linear relationship strongly depends on the clustering (e.g. for $c = 0$: $\gamma_D = -9.4\mu_T + 7$, for $c = 1$: $\gamma_D = -2.6\mu_T + 3.6$). Overskewness is in most cases associated with vertices of very high degree, and those vertices can play the role of superspreaders in models of the spread of disease or computer viruses across a network.

The cube root skewness of the degree distribution $\gamma$ is highly correlated with $\sqrt[3]{D_{\max}}$ as well as with $\log(D_{\max})$, where $D_{\max}$ is the degree of the vertex with highest degree in the network. For the networks under consideration here, the linear relationship is stronger using the logarithmic transformation, with least squares fit $\gamma_D = -1.2 + 1.4\log_{10}(D_{\max})$. The coefficient of determination $r^2 = 0.987$ indicates that $\log(D_{\max})$ and root skewness can be used interchangeably to characterize the variability in the degree distribution of networks generated by our algorithm.

### B. Clustering

Holme and Kim [32] reported a strong linear correlation between clustering probability $c$ and mean clustering coefficient $\kappa$, but the relationship shows a slight curvature which can be fitted more adequately by exponential curves of the form $\kappa = a e^c + b$, with estimates $a \in [0.25, 0.28]$ and $b \in [-0.18, -0.24]$ (Fig. 3).

The relationship between clustering and degree skewness is slightly more complicated and can best be understood by referring to the SF degree distribution. For underskewed degree distributions, increasing the clustering parameter increased the skewness of the degree distribution. For overskewed degree distributions the clustering parameter lowered the skewness of the degree distribution. Thus in all cases increasing the clustering parameter tended to make the resulting degree distribution more like the SF degree distribution, as indicated by the arrows in Fig. 2.

With this said, it is notable that the algorithm produced a variety of degree distributions with the same mean clustering, as shown in Fig. 4. To produce degree distributions with the same skewness and different mean clustering required the use of different tuning distributions.

## IV. DISCUSSION

We have exhibited an algorithm that produces contact networks with varying skewness and clustering. Investigations into real networks may demand other tuning distributions by which the degree distribution can be modified.

Using beta distributions, which can be uniform, left- or right-skewed, or even U-shaped, can provide more flexible tuning and produce a wide variety of degree distributions. Two examples, using symmetric beta distributions with expectation $\mu_T = 1/2$, are illustrated in Fig. 5. Such variations of the beta distribution can even lead to non-monotonic degree distributions. With a U-shaped beta distribution, for example, contacts are preferentially attached to vertices with lowest and highest degrees, increasing the prevalence of low-degree as well as the prevalence of high-degree vertices (superspreaders, hubs). Bell-shaped beta distributions, on the other hand, promote vertices with intermediate degree while preventing the occurrence of hubs. These examples show that beta tuning distributions should be capable of producing most degree distributions observed in real networks.

We have also shown that in a wide range of degree distributions generated by our algorithm, the skewness of the distribution is strongly correlated with the maximum degree of the network. Thus much of the descriptive information contained in the global measure (skewness) is already contained in a local measure (degree), although finding the value of the largest degree in an actual network may be very difficult.

As epidemiologists and other researchers continue to apply network techniques to a variety of areas, increased flexibility in network algorithms will be necessary. The results reported here provide a measure of that flexibility with little overhead.

## ACKNOWLEDGEMENTS

This work has been supported by EU projects SARScontrol (FP6 STREP; contract no. 003824) (HPD) and INFTRANS (FP6 STREP; contract no. 513715)) (CCL, MS), the National Science Foundation (NSF 0436298) (CCL), the MODELREL project, funded by DG SANCO (no. 2003206—SI 2378802) (MS, ME), and by the German

Ministry of Health (MS, ME). We would also like to thank the anonymous referees for their helpful comments.

**Figure 1**

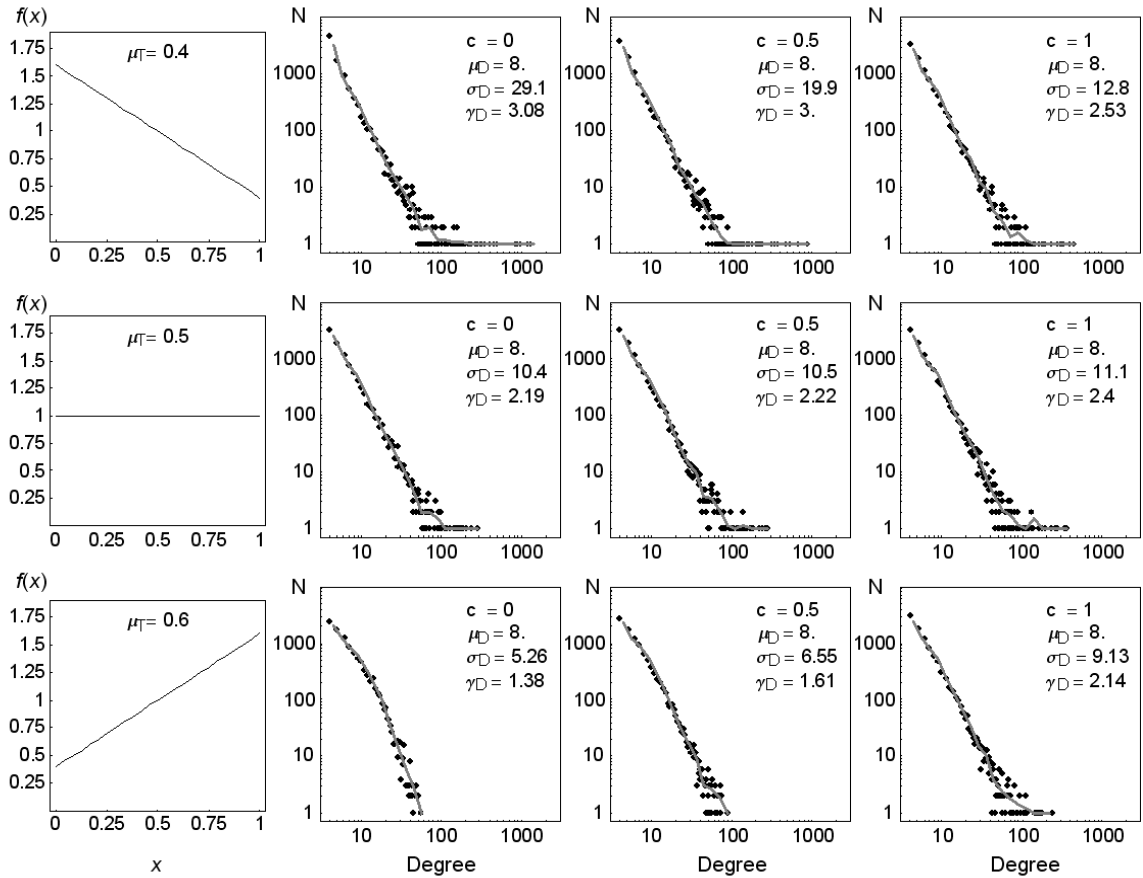

FIG. 1. Examples of degree distributions of different networks, for a subset of tuning distributions (*f(x)*, column 1) and clustering parameters $c$ (columns 2-4, with $c=0$, $c=0.5$, $c=1$). Moments for the tuning distribution (subscript $T$) and for the degree distributions (subscript $D$) are represented by $\mu$: mean, $\sigma$: standard deviation and $\gamma$: cube root of the skewness. Scalefree degree distributions with power-law behavior result from the special case $\mu_T = 1/2$, representing a uniform tuning distribution. Underskewed degree distributions result from $\mu_T > 1/2$ and overskewed degree distributions from $\mu_T < 1/2$ (see also Fig. 2). The gray line represents the average frequencies after logarithmic binning.

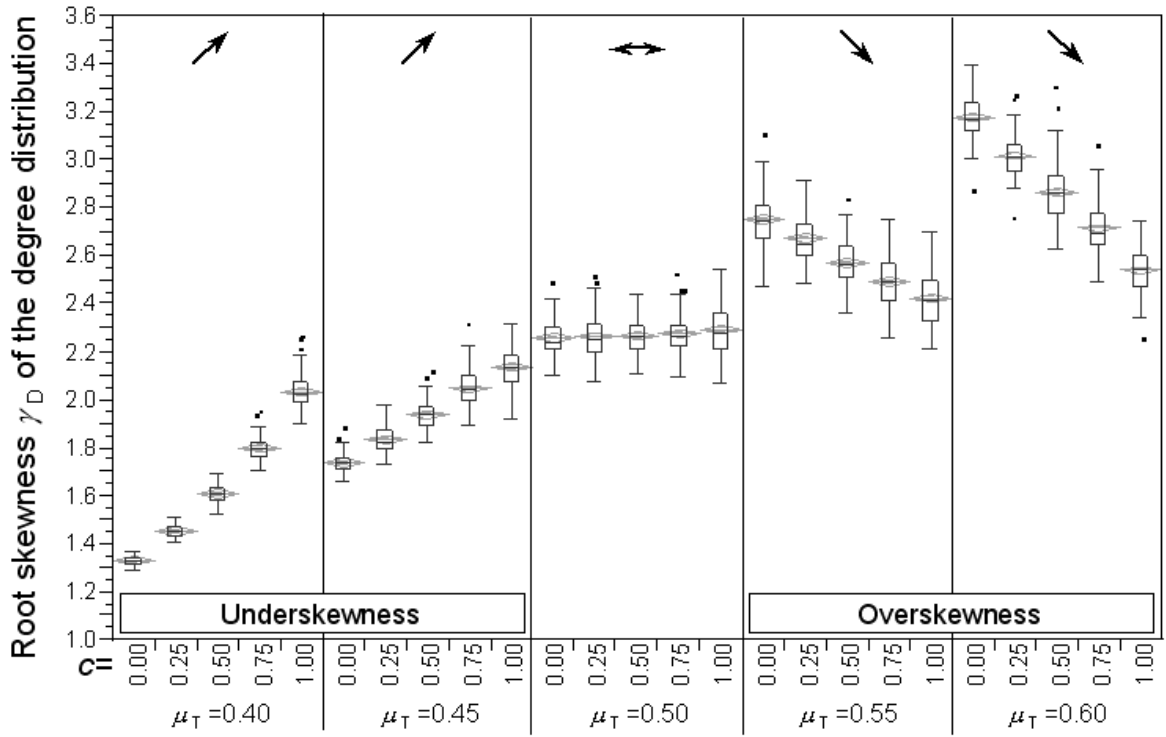

FIG. 2. Skewness distributions of different networks, grouped by the mean $\mu_T$ of the tuning distribution and clustering parameter $c$. For each of the 25 parameter combinations, 100 network realizations were simulated. The skewness distribution of the corresponding degree distributions is represented by a box & whiskers plot, diamonds (mean ± 1.5 standard error of the mean) and outliers (dots). Quartiles are represented by the lower and upper bounds of the box with the median as the line between them. Whiskers represent the quartiles ± 1.5 interquartile range and outliers are defined to lie outside this range. The skewness distributions are largely symmetric, so means and corresponding medians are close to each other. Arrows in the upper part of the graph indicate the effect of clustering on the skewness of the degree distribution (see text).

**Figure 3**

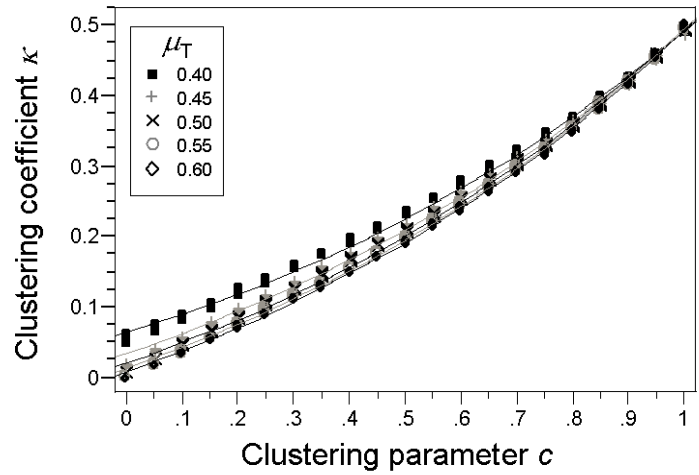

FIG. 3. Relationship between $\kappa$, the mean clustering coefficient, and $c$, the clustering parameter. For each value of $c \in \{0, 0.05, 0.1, \ldots, 1\}$ and each value of $\mu_T$, 10 networks were generated. Tuning distributions with differing expectation $\mu_T$ are indicated by differing markers as shown in the inset. Exponential least squares fits are grouped by $\mu_T$.

**Figure 4**

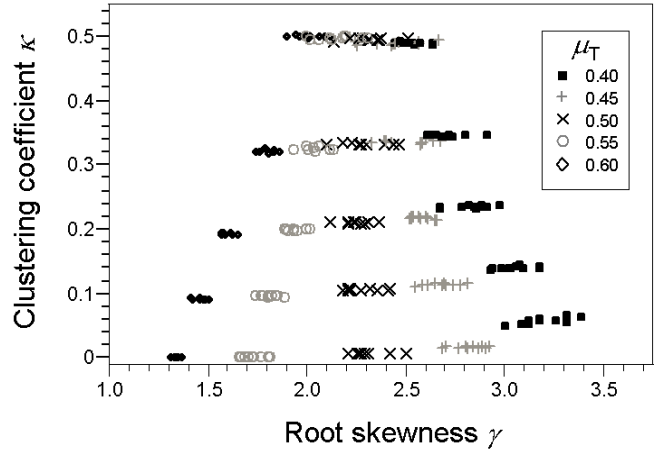

FIG. 4. Relationship between $\kappa$, the mean clustering coefficient and $\gamma$, the root skewness of the degree distribution, over 250 networks generated using the parameter values described in the main text. The horizontal bands correspond to the different values of the clustering parameter $c \in \{0, 0.25, 0.5, 0.75, 1.0\}$ (from bottom to top). Tuning distributions with differing expectation $\mu_T$ are indicated by differing markers as shown in the inset.

**Figure 5**

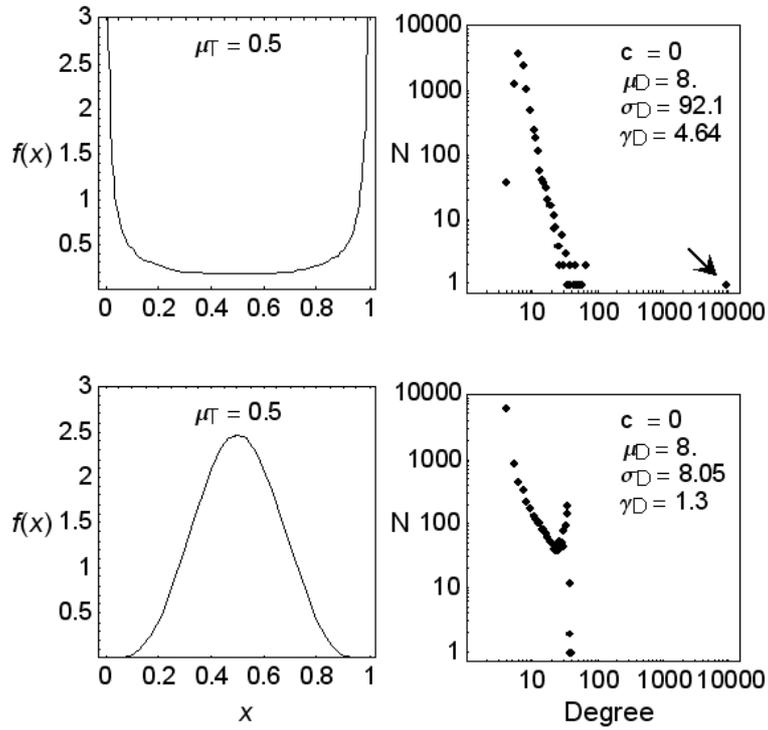

FIG. 5. Examples for degree distributions (right column) of different networks, using no extra clustering ($c=0$) and two beta distributions ($f(x)$, left column) as tuning distributions with parameters $\alpha = \beta = 0.1$ (upper row), $\alpha = \beta = 5$ (lower row). Moments for the tuning distribution (subscript $T$) and for the degree distributions (subscript $D$) are represented by $\mu$: mean, $\sigma$: standard deviation and $\gamma$: cube root of the skewness. The superspreader (hub) is indicated by an arrow.